\documentclass[twocolumn,footinbib,prb,groupedaddress]{revtex4-1}
\usepackage{amssymb}
\usepackage{amsmath}
\usepackage{graphicx}
\newcommand{\be}{\begin{equation}}
\newcommand{\ee}{\end{equation}}

\begin{document}

\title{Electronic structure modification of Si-nanocrystals with F$_4$-TCNQ}

\author{A.~Carvalho}
\email{aicarvalho@ua.pt}
\affiliation{Department of Physics, I3N, University of Aveiro, Campus Santiago, 3810-193 Aveiro, Portugal}

\author{J.~Coutinho}
\affiliation{Department of Physics, I3N, University of Aveiro, Campus Santiago, 3810-193 Aveiro, Portugal}

\author{E.~L. Silva}
\affiliation{Department of Physics, CEMDRX and CFC, Faculty of Science and Technology, University of Coimbra, Rua Larga, 3004-516 Coimbra, Portugal}

\author{S.~\"Oberg}
\affiliation{Department of Engineering Sciences and Mathematics, Lule{\aa} University of Technology, Lule{\aa}~S-97187, Sweden}

\author{M.~Rayson}
\affiliation{Department of Engineering Sciences and Mathematics, Lule{\aa} University of Technology, Lule{\aa}~S-97187, Sweden}

\author{M.~Barroso}
\affiliation{Department of Physics, I3N, University of Aveiro, Campus Santiago, 3810-193 Aveiro, Portugal}

\author{P.~R.~Briddon}
\affiliation{Electrical, Electronic and Computer Engineering, University of Newcastle upon Tyne, Newcastle upon Tyne NE1 7RU, United Kingdom}

\begin{abstract}
We use first-principles models to demonstrate how 
an organic oxidizing agent, F$_4$-TCNQ
(7,7,8,8-tetracyano-2,3,5,6-tetrafluoroquinodimethane),
modifies the electronic structure of silicon nanocrystals,
suggesting it may enhance $p$-type carrier density and mobility.
The proximity of the lowest unoccupied level of F$_4$-TCNQ to the 
highest occupied level of the
Si-nanocrystals leads to the formation of an empty hybrid state overlapping both
nanocrystal and molecule, 
reducing the excitation energy to about
0.8-1~eV in vacuum.
Hence, it is suggested that  F$_4$-TCNQ can serve both as a surface oxidant
and a mediator for hole hopping between adjacent nanocrystals.
\end{abstract}

\pacs{61.72.Bb, 61.80.Az, 71.55.Cn}
\maketitle

\section{Introduction}


Free-standing silicon nanocrystals (Si-NCs) are specially flexible building blocks 
for novel functional materials, offering the possibility of
energy band engineering from infra-red to ultra-violet,
multiple-exciton generation
and the relaxation of the momentum conservation restriction.\cite{pavesi-sinc,kovalev-prl-81-2803,beard-nl-7-2506}
These have opened the way for conceiving a range of new Si-based applications,
such as wavelength-tunable light emitters and hybrid solar cells.\cite{pavesi-sinc,liu-nl-9-449}
Silicon nanoparticles may be synthesized by gas phase methods, which
combine a high industrial upscaling production potential
with increased facility of size and surface chemistry manipulation.\cite{mangolini-am-19-2513,gupta-afm-19-696}

After recent advances, p- and n-type doping of free-standing silicon nanocrystals has been achieved, 
with P and B incorporation efficiencies close to 100\%.\cite{stegner-prl-100-026803,stegner-prb-80-165326,lechner-jap-104-053701,pi-apl-92-123102}
Given the high crystal quality of the silicon nanocrystals currently produced, the electrical performance
of the films is now predominantly
limited by the granularity and nanoscale effects.\cite{stegner-prb-80-165326}
The doping of nanostructures suffers from limitations not present in bulk silicon, 
including dopant segregation to the surface during growth,\cite{stegner-prb-80-165326} 
compensation by surface traps, self-purification and 
higher ionization energies of dopants confined in nanoparticles\cite{stegner-prb-80-165326,melnikov-prl-92-046802,iori-prb-76-085302} 
resulting from carrier confinement.\cite{pereira-prb-79-161304,chan-nl-8-596}
Further, for decreasing nanocrystal diameters, charge carrier scattering at the surface becomes more important,
and the electronic coupling between the nanoparticles becomes a dominant factor conditioning the carrier mobility.

Here, we present a strategy which may be used to improve the electrical
properties of p-type nanocrystal films.
The idea is to use an oxidizing agent to attract the electrons to the interface region between neighboring nanocrystals,
hence increasing the hole density in the nanocrystal core.
As oxidizing agent, we have chosen the organic molecule
2,3,5,6-tetrafluoro-7,7,8,8-tetracyanoquinodimethane (F$_4$-TCNQ),
used as p-type dopant in the fabrication of organic light-emitting diodes,\cite{pfeiffer-oe-4-89,walzer-CR-107-1233}
which has one of the highest electron affinities amongst organic electron acceptors
(5.08~eV).\cite{kanai-APA-95-309}
F$_4$-TCNQ has been used for $p$-type transfer doping of diamond\cite{qi-JACS-129-8084} and graphene,\cite{pinto,chen-JACS-129-10418} and to increase the work function of copper(111) crystals.\cite{romaner-PRL-99-256801}
Similarly, since its electron affinity
is larger than the ionization energy of bulk silicon, it can be used to oxidize the silicon surface.
Here we will show, using first-principles electronic structure calculations, 
that for Si nanocrystals, the position of the highest occupied electron state varies with the diameter $d$,
and becomes very close to the energy of the lowest unoccupied electron state of  F$_4$-TCNQ for $d\simeq1$~nm.
Thus, it is possible to engineer a hybrid state
shared by the nanocrystal and the molecule that
can serve as a bridging platform for hole transfer between adjacent nanocrystals.

\section{Methodology}

The electronic structure of F$_4$-TCNQ/Si-NC systems was analyzed using first-principles calculations,
carried out with a density functional code (AIMPRO).\cite{briddon-pssb-217-131,rayson-cpc-178-128} 

The local density approximation\cite{briddon-pssb-217-131} was used for the exchange and correlation energy.
Core electrons were accounted for by using the pseudopotentials of Hartwigsen {\it et al.},\cite{hartwigsen-prb-58-3641}
whereas Kohn-Sham orbitals were expanded on a 
Cartesian Gaussian basis set.\cite{goss-book}
Contracted 44G* basis functions were placed on Si and H atom sites, and 
uncontracted ddpp, dddd and ddpp basis functions were place on C, N and F sites, 
respectively (see Ref.~\onlinecite{goss-book} for a comprehensive description).

The silicon nanocrystals modeled were approximately spherical,
and had the surface passivated by hydrogen.
The diameters of the Si core ranged from $d\approx$1.5~nm (87~Si atoms) to 3.4~nm (1063~Si atoms).
For the nanocrystal total energy and electronic structure calculations, 
we used finite real space boundary conditions. 
Charge density integrations were performed using periodic boundary conditions, 
ensuring a minimum distance of 8~{\AA} between replicas of the system.
In the latter, the charge density was expanded 
in a plane wave basis set with an energy cutoff of 350~Ry.

Additionally, we investigated a composite F$_4$-TCNQ/Si-NC system,
modeled by a periodic tetragonal superlattice, with a base
of one Si-NC and a F$_4$-TCNQ molecule, having the conventional
tetragonal unit cell vectors parallel to the principle axis of the nanocrystal.
All atomic coordinates were relaxed using a conjugate gradient algorithm,
and cell coordinates were optimized using a simplex method.

\section{Results}

\subsection{Adsorption geometry}
We consider the physisorption of a single F$_4$-TCNQ molecule on the surface of the nanocrystals.
Following a previous combined theoretical and experimental study on flat (111) Si 
surfaces,\cite{furuhashi-jpcl-1-1655} 
we assume that adsorbed F$_4$-TCNQ molecules lie approximately 
parallel to the surface of a Si-NC. 
Full relaxation of the atomic coordinates using a conjugate-gradient algorithm
leaves the F$_4$-TCNQ hovering over the Si-NC curved surface,
with small deviations from its original 2D geometry. 
The hydrogen atoms at the nanocrystal surface form weak bonds 
with N and F atoms from the adsorbate.
Thus, the shortest distance between F$_4$-TCNQ and the Si-NC depends 
on the configuration of the Si-H terminations on the closest nanocrystal
facet, and on its curvature.
Equilibrium distances ranged between 2.3 and 3.2 \AA\ for the different nanocrystals studied.

\begin{figure*}[t]
\centering
\includegraphics{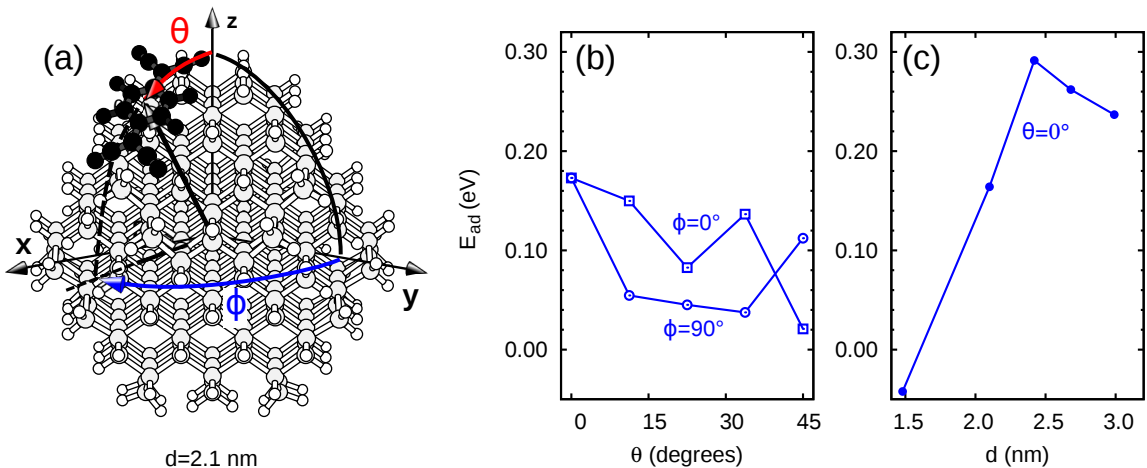}
\caption{(color online) Geometrical dependence of the adsorption energy,
$E_\mathrm{ad}$, of a F$_4$-TCNQ molecule over the surface of a Si-NC.
(a) Molecule position, represented in spherical coordinates ($R$, $\theta$ and $\phi$). 
The molecule is represented in black, whereas the Si-NC ($d=3.4$~nm)
is represented in white. (b)
Angular dependence of $E_\mathrm{ad}$ for a $d=2.1$~nm NC, and 
(c) $E_\mathrm{ad}$ for adsorption on the (001) facet as a function of NC diameter.
Lines are a guide to the eye.
\label{fig:1}}
\end{figure*}

The local geometry and density of Si-H bonds of the closest Si-NC 
surface region has an effect on the adsorption energy,
albeit with minimal impact on the electronic structure.
Different orientations of the molecule relative to the
principal axis of the nanocrystal can be generated by rotating
the molecule by $\theta$ and $\phi$ angles around the $O\hat{y}$ and 
$O\hat{z}$ axis, respectively. 
The coordinates $\theta$ and $\phi$ are defined as the angles between  
geometric center of the molecule, $\mathbf{R}$
and the $O\hat{z}$ and $O\hat{y}$ axis, respectively, 
as shown in Fig.~\ref{fig:1}(a).
For each orientation of $\mathbf{R}$ (defined by $\theta$ and $\phi$),
the molecule is placed onto the surface and all the atomic coordinates
allowed to relax. During relaxation, $\theta$ and $\phi$ changed by less than 
1$^\circ$ for all cases.

The change in the adsorption energy with respect to molecular orientation
is given in Fig.~\ref{fig:1}(b) for a $d=2.1$~nm Si-NC, for
$\phi=0^\circ$  and $\phi=90^\circ$.
The adsorption energy is here defined as
\begin{equation}
E_{\rm ad}=E_{\rm NC}+E_{\rm mol}-E_{\rm NC+mol},
\end{equation}
where $E_{\rm NC}$, $E_{\rm mol}$ and $E_{\rm NC+mol}$ are the total
energies of the isolated nanocrystal, isolated F$_4$-TCNQ molecule, 
and F$_4$-TCNQ/Si-NC system.
The highest adsorption energy is found for $\theta\simeq0^\circ$ and $\phi\simeq0^\circ$, 
when F$_4$-TCNQ lies over the (001) facet with an adsorption energy $E_\mathrm{ad}=0.18$~eV.
Thus, the molecule is placed in this position for all other calculations.
 
The adsorption energy of F$_4$-TCNQ over the (001) facet 
increases with the diameter, up to about 0.2-0.3~eV [see Fig.~\ref{fig:1}(c)].
This is higher than would be expected from the presence of weak hydrogen bonds only, as weak X-H$\cdots$Y bonds, where either  X or Y have low or moderate electronegativity, have lower energies than conventional hydrogen bonds 
(which normally have bond energies of the order of 0.2~eV\cite{kolasinski-SS-book}).
For larger nanocrystals, $E_\mathrm{ad}$ is slightly higher than for smaller nanocrystals
because in addition to the electrostatic dipole-dipole interaction associated with the
hydrogen bonds, there is an attraction resulting from the polarization
of the nanocrystal and electron transfer to the molecule, which will be
analyzed in more detail in Section~\ref{sec:den}.
This latter is absent in the smallest nanocrystal considered (1.5~nm),
and increases with nanocrystal size.
The slight decrease of the adsorption energy for 2.7~nm
and 3.0~nm nanocrystals has a different origin.
In these nanocrystals the number of SiH$_2$ di-hydride surface units
on the (001) facet is smaller, thus reducing the number of weak hydrogen bonds.

\subsection{Electronic structure and local density of states}

\begin{figure}
\includegraphics[width=8.5cm]{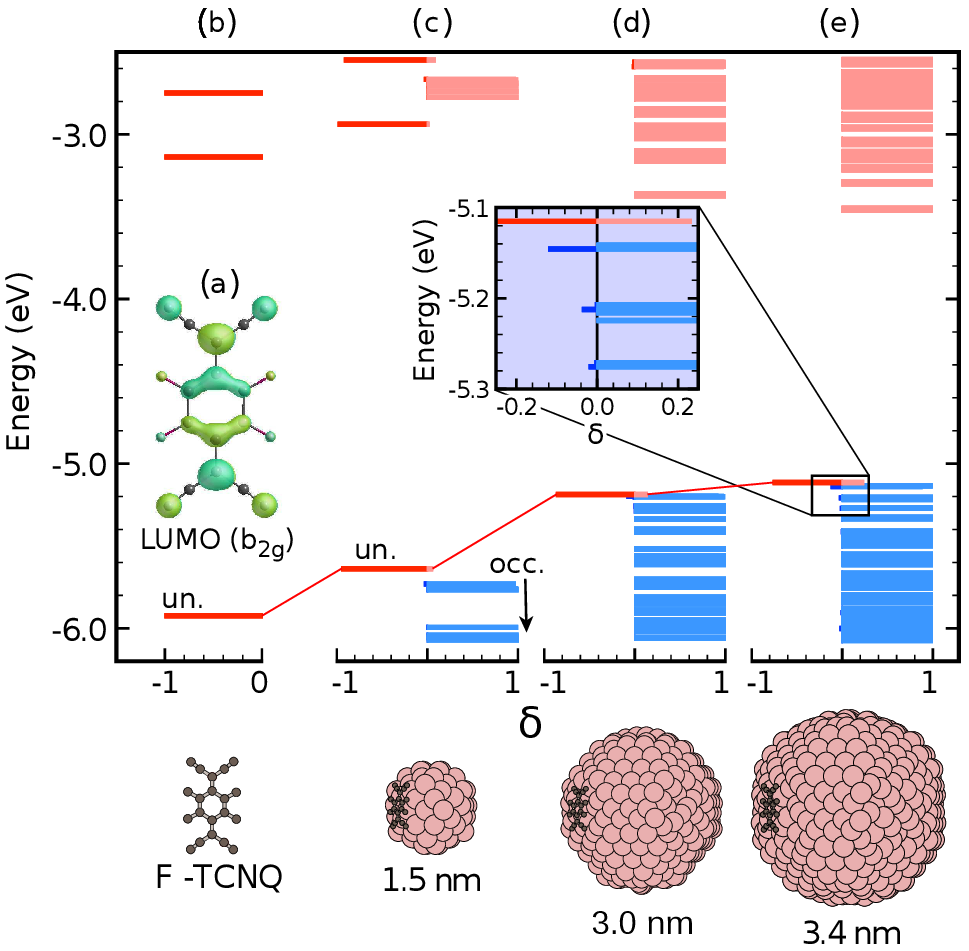}
\caption{(color online) Hybridization between F$_4$-TCNQ and Si-NC one-electron states.
(a) Isosurface plot of the F$_4$-TCNQ lowest unoccupied state ($b_{2g}$ symmetry),
and (b-e) Kohn-Sham eigenvalue energy diagrams.
The energy level diagrams are for (a) isolated F$_4$-TCNQ  
and (b-e) for F$_4$-TCNQ adsorbed on Si-NCs of increasing size.
Each state is represented by a bar of unitary length,
with the abscissa of the left and right ends of the bar indicating respectively
the relative   localization on the molecule (negative values)
and on the nanocrystal (positive values). 
Occ. and un. stand for occupied and unoccupied states, respectively.
\label{fig:2}}
\end{figure}

Inspection of the one-electron structure approximately represented,
in the density functional theory framework,
by the Kohn-Sham eigenvalues, 
and the contribution of each species to the respective eigenstates 
enlightens the changes in the electronic structure of the nanocrystals.
Populations of the Kohn-Sham states (Mulliken gross populations)
were obtained from the projection of the Kohn-Sham eigenstates on the 
basis functions.\cite{mulliken:1833}
The localization of each eigenstate ($l$) on the nanocrystal
was obtained by summing over all the basis functions localized on the 
Si and H atoms 
$\delta_l^{\rm NC}=\sum_{i\in\{NC\}}N_{li}$,
where $N_{li}$ is the population of level $l$ on basis function $i$.
To ease representation, the localization on the molecule is defined with the opposite sign
$\delta_l^{\rm mol}=-\sum_{i\in\{NC\}}N_{li}=\delta_l^{\rm NC}-1$.
The analysis for nanocrystals with $d=1.5$~nm, 3.0~nm and 3.4~nm
is represented in Fig.~\ref{fig:2},
where each horizontal line represents a Kohn-Sham state. 
The fractional localizations on the molecule ($\delta_l^{\rm mol}$) and NC ($\delta_l^{\rm NC}$)
are represented as negative and positive values of $\delta$, respectively.
For the smaller nanocrystal ($d\approx1.5$~nm),  
the interaction with a F$_4$-TCNQ adsorbate is rather weak, 
and the electronic structure of the combined system
is, in a good approximation, a superposition of the individual components.
The isolated F$_4$-TCNQ molecule has its lowest unoccupied $b_{2g}$ symmetric state $-5.92$~eV
below the vacuum level [see Fig.~\ref{fig:2}(a)].
The Kohn-Sham energy level diagram in Fig.~\ref{fig:2}(b)--(e)
shows that the lowest unoccupied Kohn-Sham (LUKS) state of the molecule lies above the 
highest occupied Kohn-Sham (HOKS) state of the nanocrystal.
The next unoccupied state of the molecule, with $b_{3u}$ symmetry,
lies within the gap for the smallest nanocrystals only.
By increasing nanocrystal diameter, however,
the HOKS energy raises, 
the LUKS energy lowers, and the density of states increases as the sharp energy levels
start forming bands. It is clear from 
Fig.~\ref{fig:2} that with the increase of nanocrystal size
the $b_{2g}$ molecular state progressively mixes with the high lying nanocrystal occupied states,
giving rise to a series of collective states that overlap both nanocrystal and F$_4$-TCNQ atoms (see inset).
For the $d=3.4$~nm nanocrystal, the population of the lowest unoccupied level of the combined system 
gives $\delta_\mathrm{NC}=0.23$, indicating that a charge is being transferred.

\subsection{Charge-transfer ratio \label{sec:den}}
\begin{figure}
\includegraphics[width=8.5cm]{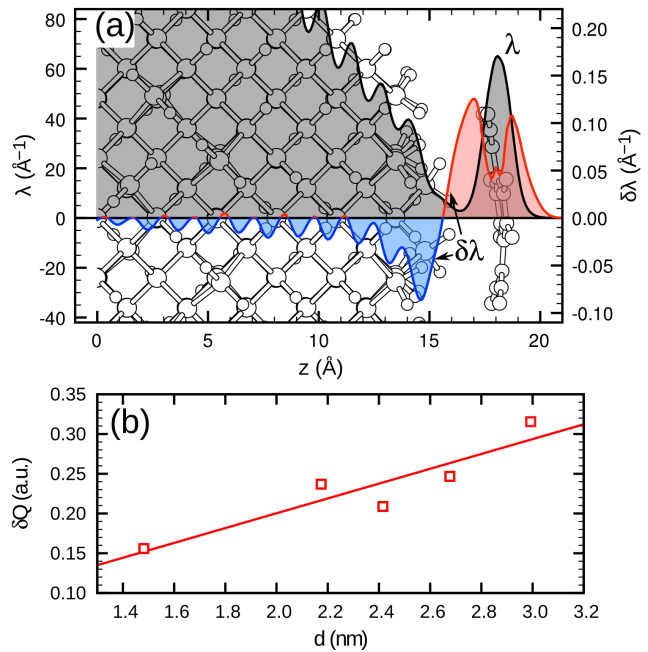}
\caption{(color online) 
Charge distribution in the F$_4$-TCNQ/Si-NC system.
(a) Integrated electron density profile $\lambda(z)$ and $\delta\lambda$
for F$_4$-TCNQ over a $d=3.0$~nm Si-NC.
The quantity $\delta\lambda$ is defined 
in Eq.~\ref{eq:lambda}.
(b) Charge transfer ratio $\delta Q$ as a function of the nanocrystal diameter. 
The line is a guide to the eye.
\label{fig:3}}
\end{figure}

The combined system
shows a displacement of the electron density ($\rho$) 
towards the the molecule.
The charge displacement is better expressed with resource to
an integrated electron density profile along the coordinate
defined by the axis connecting the center of the 
nanocrystal and the center of the molecule (here taken as $O\hat{z}$),
\be \lambda(z)=\int_0^{L_x}\int_0^{L_y}\rho(\mathbf{r})\,dx\,dy. \ee
Here $L_x$, $L_y$ and $L_z$ define the limits of the integration box,
where $\rho$ vanishes. The quantity $-e\lambda(z_0)dz$ is
the charge in the region between the planes $z=z_0$ and $z=z_0+dz$.
This is represented in Fig.~\ref{fig:3}(a) for the combined system.
We compare the integrated electron density profiles for the composed system
($\lambda_\mathrm{NC+mol}$) 
with the sum of those of the individual components ($\lambda_\mathrm{NC}$ 
and $\lambda_\mathrm{mol}$), for the same atomic coordinates. 
For all nanocrystals with $d>$1.5~nm there is a well-defined region
$z<z_m$ where 
\be \delta\lambda=\lambda_\mathrm{NC+mol}-\lambda_\mathrm{NC}-\lambda_\mathrm{mol}\label{eq:lambda}\ee
is negative (the oxidized region) and a region $z>z_m$ where $\delta\lambda$ is positive (reduced region). 
The charge displaced or transferred can be quantified by defining a charge transfer ratio
\be\delta Q=\int_{z_m}^{L_z}\delta\lambda(z)\,dz.  \ee
As depicted in Fig.~\ref{fig:3}(a), the region $z>z_m$ contains the molecule
and its closest hydrogen atoms.

The charge transfer ratio ($\delta Q$) varies from 0.16 for $d$=1.5~nm 
to 0.32 for $d=3.0$~nm.
Here the basis set superposition error is about 0.01.
As discussed in the previous sub-section, with increasing nanocrystal size,
as the position of HOKS state of the nanocrystal raises relative to 
the lowest unoccupied level of the molecule,
the mixing coefficients for the LUKS state shows an increasing contribution
of the Si-NC states, whereas the HOKS state hows an increasing contribution
of the F$_4$-TCNQ states.
Hence, $\delta Q$ increases with $d$ [Fig.~\ref{fig:3}(b)].
Further, the amount of charge transfer can be increased by adsorbing
more F$_4$-TCNQ molecules per nanocrystal.
For only three F$_4$-TCNQ molecules, the sum of the individual $\delta Q$ 
yields 0.87, already close to unity.

Moreover, the charge transfer ratio is most probably underestimated.
It is well known that the local density approximation to the exchange energy suffers 
from a self-interaction error
and favors delocalized over localized energy states. 
However, we note that the electron affinity of F$_4$-TCNQ is underestimated. 
This also happens when other approximations to the exchange and correlation energy are used
(see Ref.\footnote{
See EPAPS Document No. [number to be inserted by publisher] for comparative calculations of the electron affinity of F4-TCNQ.}
).
In contrast, the calculated work function of silicon and ionization potential of SiH$_4$
are reproduced within 0.2~eV from experiment.
Thus, the predominant source of error leads to an underestimation of the charge transfer.

\subsection{Excitation energy \label{sec:Ex}}

The excitation energy of the Si-NC with adsorbed F$_4$-TCNQ
is several times smaller than that of a bare Si-NC.
An estimate for the vertical excitation energy, $\Delta E_G^*$, may be obtained from 
total energy calculations:
\be
\Delta E_G^*=E_G^*-E_G^0,
\ee
where $E_G^0$ and $E_G^*$ are respectively the total energies of the combined system in the
ground state and
first excited state 
(obtained by promoting one electron to the lowest available empty state), 
for the ground state geometry.

The excitation energy ranges from 0.92~eV for $d$=1.5~nm to 0.65~eV for $d$=3.0~nm.
We note that the differences between LUKS and HOKS eigenvalues
are much smaller, varying between 0.09~eV for $d$=1.5~nm and 0.01~eV for $d=3.0$~nm.
Thus, the contributions to the excitation energy are the Coulomb electrostatic interaction energy, and possibly the exchange-correlation interaction energy.
In fact, according to previous theoretical models
for isolated Si-NCs, the difference between the excitonic gap and the
independent-particle gap is dominated to a large extent by classical
electrostatic contributions.\cite{delerue-JL-80-65,delerue-PRL-84-2457}
These include the interaction of the the electron and hole with their
respective image-charge distributions, and the interaction of each one of
them with the image-charge distribution of the other.

\subsection{F$_4$-TCNQ role in hole transport \label{sec:trans}}

\begin{figure}
\includegraphics{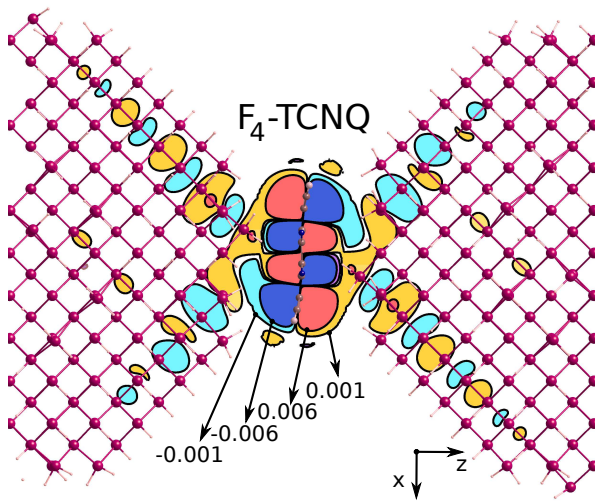}
\caption{(color online) Contour plot of the lowest unoccupied Kohn-Sham state 
for a periodic 
binary F$_4$-TCNQ/Si-NC system with $d=3.0$~nm on the $xOz$ plane.
Atoms are represented by spheres.
\label{fig:4}}
\end{figure}

We have so far considered a model system composed of a single F$_4$-TCNQ molecule
adsorbed at the surface of one Si-NC. 
As a result of the overlap between the silicon orbitals
and the orbitals of the molecule, electrons can be excited
from the occupied hybrid electron state predominantly localized on the nanocrystal to the
unoccupied hybrid electron state predominantly localized on the molecule,
becoming closer in space to a neighboring nanocrystal.
In a  nanocrystal network, this hybrid state may provide a channel 
for hole transport between neighboring nanocrystals.
To explore the nanocrystal-molecule interplay in this context, 
we used a periodic model where the nanocrystal and the molecule are repeated
along the $x$, $y$ and $z$ directions.
In this superlattice, each F$_4$-TCNQ molecule lies midway between two Si-NCs
in neighboring supercells. 
The electron density from the lowest unoccupied state of the system is depicted in Fig.~\ref{fig:4}. 
The state clearly decays into the core of both nanocrystal neighbors, supporting the
suggestion that it may provide an efficient pathway for electron transport.
Together with the sharp decrease of the excitation energy, this result suggests
that F$_4$-TCNQ may have 
an extraordinary effect on the transport properties of Si-NC films.

\section{Conclusions}

We propose that the organic acceptor F$_4$-TCNQ may be used 
to modify the electronic properties of silicon nanocrystal films.
For its large electron affinity, F$_4$-TCNQ oxidizes the nanocrystals.
The amount of charge transferred is greater the larger the nanocrystal.
In nanocrystals with $d$ close to 3~nm and larger, 
the highest occupied molecular orbitals of the nanocrystal hybridize with the $b_{2g}$ 
LUMO state of F$_4$-TCNQ to create a collective empty level.
Such communication channel is absent, for example, in B-doped
nanocrystals, where the LUMO wavefunction is confined to the nanocrystal core.
The hybrid highest occupied and lowest unoccupied states are
expected to improve long range charge transport across Si-NC networks.

\acknowledgements
We thank Prof. R. Jones for his useful suggestions.
This work was supported by FCT Portugal, 
the Calouste Gulbenkian Foundation and 
the Marie Curie Program REG/REA.P1(2010)D/22847, 
COST NanoTP, HybridSolar and Milipeia.


\end{document}